\documentclass[a4paper,12pt]{article}
\usepackage{amsmath}
\usepackage{amssymb}
\usepackage{latexsym}
\usepackage{enumerate}
\usepackage{graphicx}
\usepackage{placeins}
\usepackage{multirow}

\newcommand{\be}{\begin{eqnarray}}
\newcommand{\ee}{\end{eqnarray}}

\newcommand{\bsub}{\begin{subequations}}
\newcommand{\esub}{\end{subequations}}

\newcommand{\disfrac}[1][2]{\displaystyle\frac}
\begin{document}

\title{Localization of Energy-Momentum for a Black Hole Spacetime Geometry with
Constant Topological Euler Density}
\author{Irina Radinschi$^{\text{*1}}$, Theophanes Grammenos$^{\text{**2}}$, Farook Rahaman$^{\text{***3}}$,\\
 Andromahi Spanou$^{\text{****4}}$, Marius Mihai Cazacu$^{\text{*****1}}$,\\ Surajit Chattopadhyay$^{\text{******5}}$,
  and Antonio Pasqua$^{\text{*******6}}$\\
$^{\text{1}}$Department of Physics
``Gh. Asachi'' Technical University, \\
Iasi, 700050, Romania\\
$^{\text{2}}$Department of Civil Engineering,
University of Thessaly,\\ 383 34 Volos, Greece\\
$^{\text{3}}$Department of Mathematics, Jadavpur University,\\
Kolkata 700 032, West Bengal, India\\
$^{\text{4}}$School of Applied Mathematics and Physical Sciences,\\
National Technical University of Athens, 157 80 Athens, Greece\\
$^{\text{5}}$Department of Mathematics, Amity University,\\ Kolkata 700135, India\\
$^{\text{6}}$Department of Physics, University of Trieste,\\ 34127 Trieste, Italy\\
$^{\text{*}}$radinschi@yahoo.com, $^{\text{**}}$thgramme@civ.uth.gr,\\
$^{\text{***}}$rahaman@iucaa.ernet.in, $^{\text{****}}$aspanou@central.ntua.gr,%
\\
$^{\text{*****}}$marius.cazacu@tuiasi.ro,
$^{\text{******}}$surajcha@associates.iucaa.in,\\ $^{\text{*******}}$toto.pasqua@gmail.com}
\date{}
\maketitle

\begin{abstract}
\noindent The evaluation of the energy-momentum distribution for a
new four-dimensional, spherically symmetric, static and charged black hole spacetime geometry with constant non-zero topological Euler density is performed by using the energy-momentum complexes of Einstein and M\o ller. This black hole solution was recently developed in the context of the coupled Einstein--non-linear electrodynamics of the Born-Infeld type. The energy is found to depend on the
mass $M$ and the charge $q$ of the black hole, the cosmological constant $\Lambda$ and the radial coordinate $r$, while in both prescriptions all the
momenta vanish. Some limiting and particular cases are analyzed and discussed, illustrating the rather extraordinary character of the spacetime geometry considered.
\end{abstract}

\maketitle

\textbf{Keywords}: Energy-Momentum Complexes, Black Holes,
Topological Euler Density \newline
\textit{PACS Numbers}: 04.20.-q, 04.20.Cv, 04.70.Bw\newline
\textit{MSC}: 83C57, 83C40, 58K65

\section{Introduction}

The issue of energy-momentum localization systematised researchers' work in a special way. Looking deeply into the problem, it was clear that the main difficulty consists in the lack of a proper definition for the energy density of gravitational backgrounds. In this light, much research work has been done over the last years concerning the best tools used for the energy-momentum localization. A brief survey points out the leading role played notably by super-energy tensors \cite{Bel}-\cite{Senovilla}, quasi-local expressions \cite{Brown}-\cite{Szabados} and the famous energy-momentum complexes of Einstein \cite{Einstein}-\cite{Trautman}, Landau-Lifshitz \cite{Landau}, Papapetrou \cite{Papapetrou}, Bergmann-Thomson \cite{Bergmann}, M\o ller \cite{Moller} and Weinberg \cite{Weinberg}. Among the aforementioned computational tools, the energy-momentum complexes have been proven to be interesting and useful as well due to the diverse and numerous reasonable expressions that can be obtained by their application. Some observations are in order here. First, according to their underlying mathematical mechanism, their construction involves the use of two parts, one for the matter and one for the gravitational field. Second, despite the fact that the energy-momentum complexes allow one to obtain many interesting and physically meaningful results for different space-time geometries, their construction is connected to an inherent central problem, namely their coordinate dependence. As it is well-known from the relevant literature, this problem has found a solution in the case of the M\o ller energy-momentum complex. Indeed, the calculations for the energy-momentum of a given gravitational background in the M\o ller prescription enable the use of any coordinates such that the energy density component transforms as a four-vector density under purely spatial coordinate transformations for metrics with a line element of the form $ds^2=g_{00}dt^2-g_{ij}dx^idx^j$. As for the other energy-momentum complexes, the Schwarzschild Cartesian coordinates and the Kerr-Schild Cartesian coordinates have to be utilised for the calculations providing physically reasonable results for the cases of space-time geometries in $(3+1)$, $(2+1)$ and $(1+1)$ dimensions (see, e.g., \cite{Ayon}-\cite{Mat} and references therein). At this point it should be noticed that, in order to avoid
the coordinate dependence, an alternative method for the computation of the energy and momentum distributions is provided by the teleparallel equivalent of general relativity and certain modified versions of the teleparallel theory (see, e.g., \cite{Maluf}-\cite{daSilva} and references therein).

Regarding the Einstein, Landau-Lifshitz, Papapetrou, Bergmann-Thomson, Weinberg and M\o ller energy-momentum complexes there is an agreement with the definition of the quasi-local mass introduced by Penrose \cite{Penrose} and developed by Tod \cite{Tod} for some gravitational backgrounds. We point out that some rather recent works show that several energy-momentum complexes ``provide the same results'' for any metric
of the Kerr-Schild class and indeed even for solutions that are more general than those of the Kerr-Schild class (see, e.g., \cite{Aguirregabiria} and \cite{Xulu_2}, and
the interesting article  \cite{Virb_1} on the subject). Further, the entire historical development of the energy-momentum complexes that started with the formulation of their definitions also includes the attempts made for their rehabilitation \cite{Chang}-\cite{Sun_Nester}. In this sense, perhaps the most interesting issue was the fact that different energy-momentum complexes yield the same results for the energy-momentum distribution in the case of various gravitating systems.

The present paper has the following structure: in Section 2 we describe the new class of four-dimensional spherically symmetric, static and charged black hole solutions with constant non-zero topological Euler density which we will consider. Section 3 focuses on the presentation of the Einstein and M\o ller prescriptions used for
performing the calculations. In Section 4 we present the calculations and the results obtained for the energy and momentum distributions. Finally, in the Discussion provided in Section 5, we give a brief description of the results obtained as well as some limiting and particular cases. Throughout we use geometrized units ($c=G=1$) and the signature  chosen  is $(+,-,-,-)$. Further, the calculations are performed by using the
Schwarzschild Cartesian coordinates $\{t, x, y, z\}$ for the Einstein energy-momentum complex and the Schwarzschild coordinates $\{t, r, \theta, \varphi \}$ for the M\o ller energy-momentum complex. Finally, Greek indices run from $0$ to $3$, while Latin indices range from $1$ to $3$.

\section{Description of the New Black Hole Solution with Constant Topological Euler Density}

This section deals with the presentation of the new four-dimensional spherically symmetric, static and charged black hole solution with constant topological Euler density \cite{Bargueno} examined in the present study. Connecting geometry with topology, from the generalised Gauss--Bonnet theorem (see, e.g. \cite{Gilkey}) applied to four dimensions the Euler--Poincar\'e characteristic is obtained by the integral of the Euler density 
\begin{equation}\label{Euler_density}
{\mathcal G}=\frac{1}{32\pi ^{2}}(R^{\kappa\lambda\mu \nu}R_{\kappa\lambda\mu \nu}
-4R^{\kappa\lambda }R_{\kappa\lambda}+R^{2}),
\end{equation}
where $R_{\kappa\lambda\mu \nu}$ is the Riemann curvature tensor, $R_{\kappa\lambda}$ is the Ricci tensor, and $R$ is the Ricci scalar\footnote{Often the topological Euler density is given without the factor $1/32\pi^2$ (see, e.g. \cite{Ammon}). In fact, the terms in the parenthesis constitute the so-called ``quadratic Gauss-Bonnet term'' in the Lovelock gravity Lagrangian.}. For a general, spherically symmetric and static geometry described by the line element 
\begin{equation}\label{line_element}
ds^{2}=f(r)dt^{2}-f(r)^{-1}dr^{2}-r^{2}(d\theta ^{2}+\sin ^{2}\theta
d\varphi ^{2}) 
\end{equation}
eq.~(\ref{Euler_density}) becomes 
\begin{equation}\label{G}
{\mathcal G}=\frac{4}{r^{2}}\{f^{\prime }(r)^{2}+[f(r)-1]f^{\prime \prime }(r)\},
\end{equation}
where the constant $32\pi^2$ has been absorbed in ${\mathcal G}$. For constant topological density ${\mathcal G}=\alpha \neq 0$, Eq. (\ref{G}) gives for the metric function
\begin{equation}\label{metric_function}
f(r)=1\pm \left(1-2A+Br+\frac{\alpha r^{4}}{24}\right)^{1/2}, 
\end{equation}
with $A$, $B$ arbitrary constants. In what follows, we will keep only the negative sign of Eq.~(\ref{metric_function}), as it is the one leading to black hole solutions.

The new solution derived in \cite{Bargueno} is based on the coupling of gravity to non-linear electromagnetic fields as described by the non-linear generalisation of Maxwell's electrodynamics according to the Born--Infeld theory. Thus, in the chosen case of electrovacuum, the radial electric field
\begin{equation}\label{electric_field}
E(r)= \frac{r^{2}}{4q}(4R^{\mu \nu }R_{\mu \nu }-R^{2})^{1/2}
\end{equation}
where $q$ the electric charge, solves the Einstein--non-linear electrodynamics coupled system with the Ricci tensor $R_{\mu\nu}$ and the Ricci scalar $R$ calculated by using the line element (\ref{line_element}) and the metric function (\ref{metric_function}). In fact, with the values $A=\frac{1}{2}+\frac{q^{2}\Lambda }{3}$, $B=\frac{4M\Lambda}{3}$ and 
$\alpha=\frac{8\Lambda^{2}}{3}$, with $\Lambda$ is the cosmological constant and $M$ is the mass of the black hole, the line element (\ref{line_element}) with the metric function (\ref{metric_function}) becomes
\begin{equation}\label{final_line_element}
\begin{split}
ds^{2}= & \left[1-\sqrt{\frac{4M\Lambda r}{3}+\frac{\Lambda^{2}r^{4}}{9}- \frac{2q^{2}\Lambda}{3}}\right]dt^{2}\\
&-\left[1-\sqrt{\frac{4M\Lambda r}{3}+\frac{\Lambda^{2}r^{4}}{9}- \frac{2q^{2}\Lambda}{3}}\right]^{-1}dr^{2}
-r^{2}(d\theta ^{2}+\sin^{2}\theta \,d\varphi^{2}),
\end{split}
\end{equation}
describing a Reissner--Nordstr\"{o}m--de Sitter black hole spacetime geometry. 

Now one can distinguish between two different cases, namely the massive case ($M\neq 0$) and the massless case ($M=0$). In the first case, Eq.~(\ref{final_line_element}), when $q=0$ and $\Lambda>0$, shows that the geometry is regular everywhere except at the origin $r=0$. However, this case has no particular interest from the electrodynamic viewpoint.  
Black hole solutions with zero mass were proposed as a conjecture by A. Strominger \cite{Strominger} in order to explain conifold singularities in the context of string theory. In particular, the ten-dimensional IIA (resp. IIB) string theory admits black D2- (resp. D3-) brane solutions with a mass proportional to their area. After applying a Calabi-Yau compactification these solutions may wrap around minimal 2-surfaces (resp. 3-surfaces) in the Calabi-Yau space and they appear as four-dimensional black holes. As the area of the surface around which they wrap is let to go to zero, the corresponding extremal black holes become massless, topologically stable, structures. In fact, the existence of stable black hole solutions with zero ADM mass was shown in \cite{Behrndt} although their relation to the massless solutions suggested in \cite{Strominger} is still not clarified. Indeed, since then there has been an increasing interest in massless black hole solutions (see, e.g. \cite{Emparan}-\cite{Sudarsky} and references therein). Recently, massless black hole solutions were obtained from the dyonic black hole solution of the Einstein--Maxwell--dilaton thery \cite{Goulart}. Although the black hole solution examined here does not originate from string theory we will proceed to the consideration of the massless ($M=0$) case despite its physically dubious character. 

Thus for  $M=0$, the line element (\ref{final_line_element}) becomes
\begin{equation}\label{zero_line_element}
ds^{2}=  \left[1-\sqrt{\frac{\Lambda^{2}r^{4}}{9}- \frac{2q^{2}\Lambda}{3}}\right]dt^{2}
-\left[1-\sqrt{\frac{\Lambda^{2}r^{4}}{9}- \frac{2q^{2}\Lambda}{3}}\right]^{-1}dr^{2}
-r^{2}(d\theta ^{2}+\sin^{2}\theta \,d\varphi^{2}).
\end{equation}
Here, when $q=0$ and $\Lambda\neq 0$ the de Sitter solution is obtained, while for $q\neq 0$ but $\Lambda=0$ one gets the Minkowski solution. In the case $q\neq 0$ and $\Lambda <0$, an event horizon exists and the spacetime is singular at the origin $r=0$, while the electric field is everywhere regularisable. An overall detailed study of the black hole's behaviour in the massive as well as in the massless case is presented in \cite{Bargueno}.

\section{Einstein and \ M\o ller Energy-Momentum Complexes}

In this section we outline the definitions of the Einstein and M\o ller
energy-momentum complexes.

The expression for the Einstein energy-momentum complex \cite{Einstein} in the case of a
(3+1)-dimensional gravitational background was later found to be given by 
\begin{equation}\label{Einstein_complex}
\theta _{\nu }^{\mu }=\frac{1}{16\pi }h_{\nu ,\,\lambda }^{\mu \lambda }. 
\end{equation}
The Freud superpotentials $h_{\nu }^{\mu \lambda }$ in (\ref{Einstein_complex}) are calculated by the compact formula (found by Landau and Lifshitz)
\begin{equation}\label{von_Freud_super}
h_{\nu }^{\mu \lambda }=\frac{1}{\sqrt{-g}}g_{\nu \sigma }\left[ -g(g^{\mu
\sigma }g^{\lambda \kappa }-g^{\lambda \sigma }g^{\mu \kappa })\right]
_{,\kappa }
\end{equation}%
and satisfy the antisymmetric property 
\begin{equation}\label{E_anti}
h_{\nu }^{\mu \lambda }=-h_{\nu }^{\lambda \mu }. 
\end{equation}%
We notice that in the Einstein prescription the local conservation law is
respected:
\begin{equation}\label{E_conserv}
\theta _{\nu ,\,\mu }^{\mu }=0.
\end{equation}
Consequently, the energy and momentum can be calculated in Einstein's prescription by 
\begin{equation}\label{E_em}
P_{\nu }=\iiint \theta _{\nu }^{0}\,dx^{1}dx^{2}dx^{3}.
\end{equation}
Here, $\theta _{0}^{0}$ and $\theta _{i}^{0}$ represent the energy and
momentum density components, respectively.

Applying Gauss' theorem, the energy-momentum reads
\begin{equation}\label{EG_em}
P_{\mu }=\frac{1}{16\pi }\iint  h_{\mu }^{0i}n_{i}dS,
\end{equation}
with $n_{i}$ the outward unit normal vector over the surface $dS.$ In Eq.~(\ref{EG_em}) the component $P_{0}$ represents the energy.

According to \cite{Moller},  the M{\o }ller energy-momentum complex is
\begin{equation}\label{Moller_cmplex}
\mathcal{J}_{\nu }^{\mu }=\frac{1}{8\pi }M_{\nu \,\,,\,\lambda }^{\mu
\lambda }, 
\end{equation}%
with the M{\o }ller superpotentials $M_{\nu }^{\mu \lambda }$ given by 
\begin{equation}\label{Moller_super}
M_{\nu }^{\mu \lambda }=\sqrt{-g}\left( \frac{\partial g_{\nu \sigma }}{%
\partial x^{\kappa }}-\frac{\partial g_{\nu \kappa }}{\partial x^{\sigma }}%
\right) g^{\mu \kappa }g^{\lambda \sigma }. 
\end{equation}%
The M{\o }ller superpotentials $M_{\nu }^{\mu \lambda }$ satisfy the antisymmetric property 
\begin{equation}\label{M_anti}
M_{\nu }^{\mu \lambda }=-M_{\nu }^{\lambda \mu }. 
\end{equation}%
As in the case of the Einstein prescription, in the M{\o }ller prescription
the local conservation law is also satisfied:
\begin{equation}\label{M_conserv}
\frac{\partial \mathcal{J}_{\nu }^{\mu }}{\partial x^{\mu }}=0.
\end{equation}%
In (\ref{M_conserv}) the component $\mathcal{J}_{0}^{0}$ represents the energy density and the  $\mathcal{J} _{i}^{0}$ give the momentum density components.

For the M\o ller energy-momentum complex, the energy-momentum
distributions are given by 
\begin{equation}\label{M_em}
P_{\nu }=\iiint \mathcal{J}_{\nu }^{0}dx^{1}dx^{2}dx^{3}. 
\end{equation}%
In particular, the energy distribution can be computed by 
\begin{equation}\label{M_e}
E=\iiint \mathcal{J}_{0}^{0}dx^{1}dx^{2}dx^{3}.
\end{equation}%
Using Gauss' theorem one gets
\begin{equation}\label{MG_em}
P_{\nu }=\frac{1}{8\pi }\iint M_{\nu }^{0i}n_{i}dS,
\end{equation}
where, again, $n_i$ is the outward unit normal vector over the surface $dS$.

\section{Energy and Momentum Distribution for the Black Hole Solution
with Constant Topological Euler Density}

To compute the energy and momenta with the Einstein energy-momentum complex,
we have to transform the metric given by the line element (\ref{final_line_element}) in
Schwarzschild Cartesian coordinates by using the coordinate transformation $%
x=r\sin \theta \cos \varphi$, $y=r\sin \theta \sin \varphi$, $z=r\cos \theta$. Thus, we obtain a new form for the line element:
\begin{equation}\label{new_line_element}
ds^{2}=f(r)dt^{2}-(dx^{2}+dy^{2}+dz^{2})-\frac{f^{-1}(r)-1}{r^{2}}
(xdx+ydy+zdz)^{2}.
\end{equation}
In Schwarzschild Cartesian coordinates for $\nu =0,1,2,3$ and $i=1,2,3$ we
find the following vanishing components of the superpotentials $h_{\nu }^{0i}$:
\begin{equation}\label{E_super_comp}
\begin{split}
h_{1}^{01} &= h_{1}^{02}=h_{1}^{03}=0, \\
h_{2}^{01} &= h_{2}^{02}=h_{2}^{03}=0,\\
h_{3}^{01} &= h_{3}^{02}=h_{3}^{03}=0. 
\end{split}
\end{equation}
In order to compute the non-vanishing components of the superpotentials in
the Einstein prescription we use (\ref{von_Freud_super}) and we obtain the following expressions:
\begin{equation}\label{E_super_1}
h_{0}^{01}=\frac{2x}{r^{2}}\sqrt{\frac{4M\Lambda r}{3}+\frac{\Lambda
^{2}r^{4}}{9}-\frac{2\,q^{2}\,\Lambda }{3}}, 
\end{equation}
\begin{equation}\label{E_super_2}
h_{0}^{02}=\frac{2y}{r^{2}}\sqrt{\frac{4M\Lambda r}{3}+\frac{\Lambda
^{2}r^{4}}{9}-\frac{2q^{2}\Lambda }{3}}, 
\end{equation}
\begin{equation}\label{E_super_3}
h_{0}^{03}=\frac{2z}{r^{2}}\sqrt{\frac{4M\Lambda r}{3}+\frac{\Lambda
^{2}r^{4}}{9}-\frac{2q^{2}\Lambda }{3}}. 
\end{equation}
With the aid of the line element (\ref{new_line_element}), the expression for the energy given
by (\ref{EG_em}) and the expressions (\ref{E_super_1})-(\ref{E_super_3}) for the superpotentials, we get the
energy distribution for the examined black hole in the Einstein prescription:
\begin{equation}\label{Einstein_energy}
E_{E}=\frac{r}{2}\sqrt{\frac{4M\Lambda r}{3}+\frac{\Lambda ^{2}r^{4}}{9}%
-\frac{2q^{2}\Lambda }{3}}.  
\end{equation}
In order to calculate the momentum components we employ (\ref{EG_em}) and (\ref{E_super_comp}) and
performing the calculations we find that all the momenta vanish:
\begin{equation}
P_{x}=P_{y}=P_{z}=0. 
\end{equation}

In the M\o ller prescription we perform the calculations in Schwarzschild
coordinates $\{t, r,\theta, \varphi \}$ with the aid of the line element (\ref{final_line_element}) and we find only one non-vanishing superpotential:
\begin{equation}\label{Moller_super_1}
M_{0}^{01}=-\frac{1}{6}\frac{12\,M\,\Lambda +4\,\Lambda ^{2}\,r^{3}}{\sqrt{%
12\,M\,\Lambda \,r+\Lambda ^{2}\,r^{4}-6\,q^{2}\,\Lambda }}r^{2}\,\sin
\theta.
\end{equation}%
Using the above expression for the superpotential and the
expression for the energy obtained from (\ref{MG_em}), we get the energy in the M\o ller
prescription:
\begin{equation}\label{Moller_energy}
E_{M}=-\frac{r^{2}}{12}\frac{12M\Lambda +4\Lambda ^{2}r^{3}}{\sqrt{%
12M\Lambda \,r+\Lambda ^{2}r^{4}-6q^{2}\Lambda }}.
\end{equation}
Finally, all the momenta are found to be zero:
\begin{equation}
P_{r}=P_{\theta }=P_{\varphi }=0.
\end{equation}

\section{Discussion}

Our paper focuses on the analysis of the energy-momentum localization for a new four-dimensional,
spherically symmetric, static and charged black hole spacetime geometry with constant
non-zero topological Euler density, given by the line element (\ref{final_line_element}). The solution describes a Reissner--Nordstr\"om--de Sitter spacetime geometry as the result of the coupling of Einstein gravity with non-linear electrodynamics of the Born--Infeld type. For $q=0$ the solution has a near-de Sitter behavior, while for $\Lambda >0$  the solution is regular everywhere except at the origin $r=0$. To perform our study we use the Einstein and M\o %
ller energy-momentum complexes. The calculations provide the well-defined
expressions (\ref{Einstein_energy}) and (\ref{Moller_energy}) for the energy distribution in both prescriptions. These energy
distributions depend on the mass $M$ and the charge $q$ of the black hole,
the cosmological constant $\Lambda$ and the radial coordinate $r$, while in
both energy-momentum complexes all the momenta vanish.

In order to study the limiting behavior of the energy distributions
obtained by the Einstein and M\o ller prescriptions, we consider the energy for $r\rightarrow \infty$ in the uncharged case  $q=0$ for the massive ($M \neq 0$) black hole, and for  $r\rightarrow \infty $, $\Lambda
=0$ and $q=0$ for the massless  ($M=0$) black hole.

Starting with the massive black hole ($M\neq 0$) the results for the
limiting cases $r\rightarrow \infty $ and $q=0$ are
presented in Table 1.
\begin{table}[h!]
\centering
\begin{tabular}{|c|c|c|}
\hline
Energy & $r\rightarrow \infty $ & $q=0$ \\ \hline
&&\\
$E_{E}$ & $\infty $  & $\displaystyle{\frac{r}{2}\sqrt{\frac{4M\Lambda
r}{3}+\frac{\Lambda^{2}r^{4}}{9}}}$ \\ 
&&\\
$E_{M}$ & $-\,\infty $ &  $-\displaystyle{\frac{r^{2}}{12}\frac{%
12M\Lambda +4\Lambda^{2}r^{3}}{\sqrt{12M\Lambda r+\Lambda^{2}r^{4}}}}$ \\
&&\\
 \hline
\end{tabular}%
\caption{Limiting behaviour of the energy of the massive ($M\neq 0$) black hole solution in the Einstein and M\o ller prescriptions.}
\end{table}

For the charged ($q\neq 0$) black hole without cosmological constant ($\Lambda=0$), the spacetime geometry becomes the Minkowski geometry. If the black hole is uncharged ($q=0$) and the cosmological constant is non-zero ($\Lambda >0$), then the spacetime is regular everywhere except at the origin $r=0$, as it is inferred from the calculation of the curvature invariants (Kretschmann and Ricci) in \cite{Bargueno}. Indeed, the fact that, despite the regular behavior of the metric, the energy diverges at infinity in both prescriptions, can possibly be attributed to the de Sitter like asymptotic behavior of the solution. The evolution of the Einstein energy and of the M\o ller energy with respect to the $r$ coordinate, as given by Eqns. (\ref{Einstein_energy}) and (\ref{Moller_energy}), is presented for $\Lambda > 0$ and $\Lambda < 0$ in Fig.~1. In fact, the M\o ller energy given by Eq.~(\ref{Moller_energy}) is negative for a large range of values for $r$ and the parameters $M$, $\Lambda$, and $q$. The observed negativity of the energy might sound a note of warning regarding the physical interpretation of this result and, consequently, the merits of the M\o ller complex despite the physically acceptable results obtained by the latter in many other cases. Furthermore, if one considers the behavior of the energy regarding $r$, one concludes that the fourth degree polynomial $12M\Lambda \,r+\Lambda^{2}r^{4}-6q^{2}\Lambda$ in the denominator of Eq.~(\ref{Moller_energy}) vanishes for several sets of values for the parameters $M$, $\Lambda$, and $q$ with its roots consisting in two real and two complex conjugate solutions. As a result, at finite distances from the black hole the Einstein energy, given by Eq.~(\ref{Einstein_energy}), vanishes, while the M\o ller energy, given by Eq.~(\ref{Moller_energy}), diverges. Equally interesting is the vanishing of the third degree polynomial $12M\Lambda +4\Lambda^{2}r^{3}$ in the nominator of Eq.~(\ref{Moller_energy}). In this case, one real and two complex conjugate solutions are obtained for every finite value of the mass parameter $M$ and $\Lambda <0$. As a result, the M\o ller energy vanishes at finite distances from the black hole. We have not found any plausible explanation for this rather pathological behavior of the energy in the two prescriptions, beside the fact that the black hole solution considered is quite peculiar. Some more light could be shed on this strange state of things through the comparison of the present results with the energy calculated, from example, by the Landau-Lifshitz and the Weinberg energy-momentum complexes for the metric considered, a task left for future work.

\begin{figure}[thbp]
\begin{center}
\begin{tabular}{rl}
\includegraphics[width=7.cm]{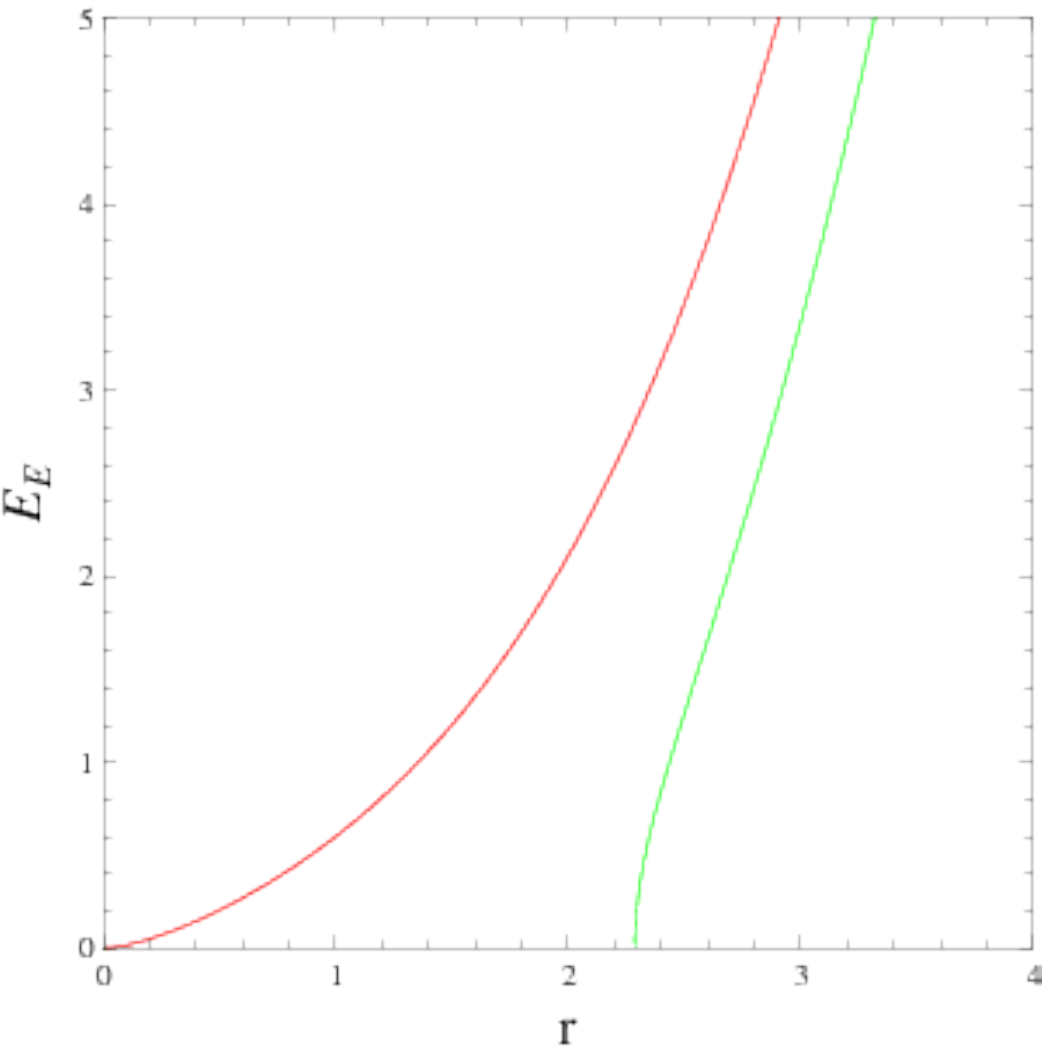}&
\includegraphics[width=7.43cm]{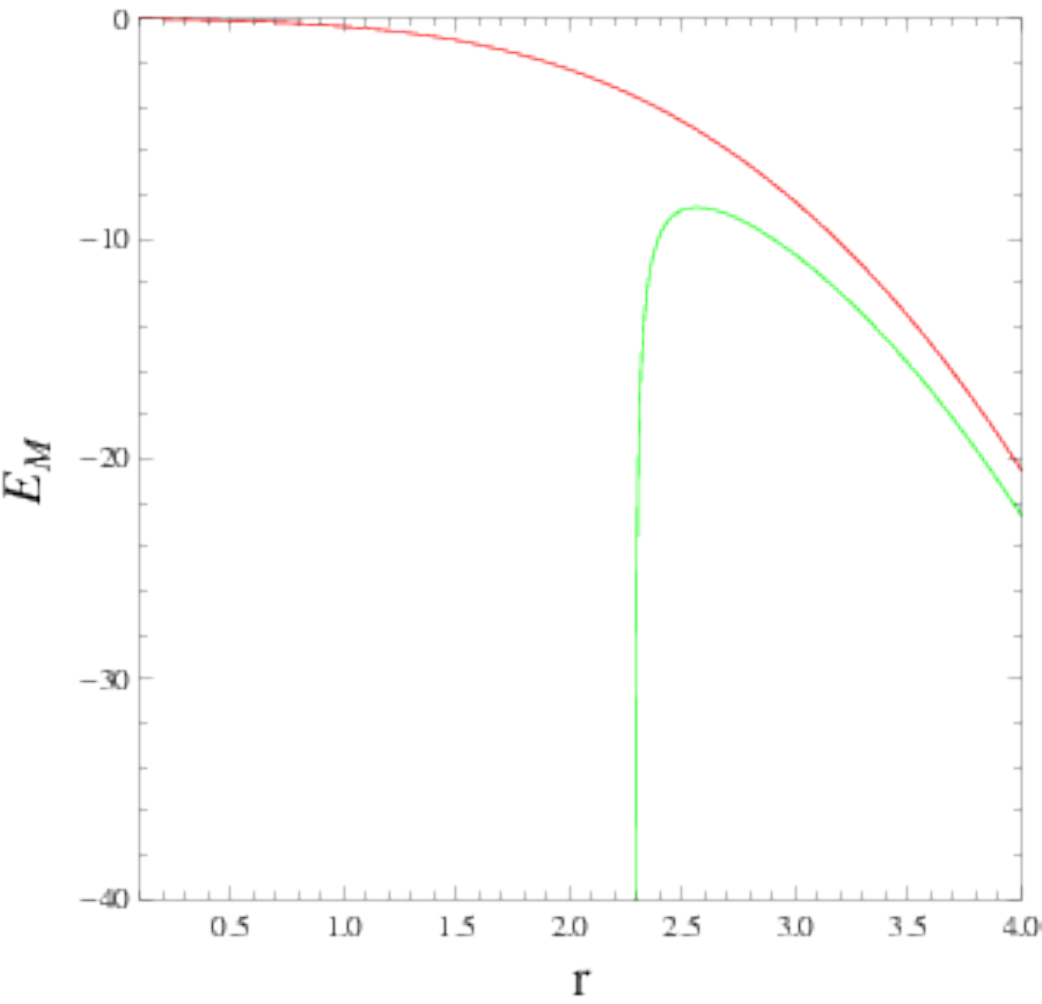}
\end{tabular}
\caption{Evolution of the Einstein  energy $E_E$ (left) and of the M\o ller energy $E_M$ (right) with respect to  the $r$ coordinate for the massive and charged black hole. In both cases we have chosen $M=1$ and $q=1$, while the red line and the green line correspond to $\Lambda=1$ and $\Lambda=-1$, respectively.}
\end{center}
\vspace{-.5cm}
\end{figure}

In the massless case ($M=0$), when the black hole is charged  ($q\neq 0$) and the cosmological constant is non-zero ($\Lambda \neq 0$), the spacetime geometry is described by the line element (\ref{zero_line_element}). In fact, an event horizon appears at $r_{\scriptsize\text{eh}}=[\frac{9}{\Lambda^{2}}$ $+\frac{6q^{2}}{\Lambda }]^{\frac{1}{4}}$ for a negative cosmological constant $\Lambda<-3/(2q^2)$. Further, when $\Lambda>0$, at the position $r_s=(\frac{6q^2}{\Lambda})^{1/4}$ a singularity appears, while, when $\Lambda<0$, this singularity can be avoided but there appears another singularity at $r=0$  (see \cite{Bargueno}). If this massless black hole has no charge ($q=0$) but the cosmological constant in non-zero ($\Lambda \neq 0$), then a de Sitter spacetime geometry is obtained:
\begin{equation}
ds^{2}=\left(1-\sqrt{\frac{\Lambda^{2}r^{4}}{9}}\right)dt^{2}-\left(1-\sqrt{\frac{\Lambda^{2}r^{4}}{9}}\right)^{-1}dr^{2}-r^{2}(d\theta ^{2}+\sin ^{2}\theta d\varphi^{2}),
\end{equation}
with a cosmological horizon appearing at $r_{\scriptsize\text{ch}}=\sqrt{\frac{3}{\Lambda }}$. The energy of the massless black hole for the charged and the uncharged case, computed in the Einstein and M\o ller prescriptions, is presented in Table 2.

\begin{table}[h!]
\centering
\begin{tabular}{|c|c|c|}
\hline
Energy & $ q\neq 0, \Lambda \neq  0$ & $q=0, \Lambda \neq 0$ \\ 
\hline
&&\\
$E_{E}$ & $\displaystyle{\frac{r}{2}\sqrt{\frac{\Lambda^{2}r^{4}}{9}-\frac{2q^{2}\Lambda}{3}}}$ & $\displaystyle{\frac{r}{2}\sqrt{\frac{\Lambda^{2}r^{4}}{9}}}$ \\ 
&&\\
$E_{M}$ & $-\displaystyle{\frac{1}{3}\frac{\Lambda^{2}r^{5}}{\sqrt{\Lambda
^{2}r^{4}-6q^{2}\Lambda}}}$ & $-\displaystyle{\frac{1}{3}\frac{\Lambda^{2}r^{5}}{\sqrt{\Lambda^{2}r^{4}}}}$ \\
&&\\
 \hline
\end{tabular}%
\caption{Energy of the massless ($M=0$) black hole solution in the Einstein and M\o ller prescriptions.}
\end{table}

In Table 3 we present the limiting behaviour of the Einstein energy and the M\o ller energy for the charged and the uncharged massless ($M=0$) black hole as $r\rightarrow \infty$. In both cases the cosmological constant is non-zero.

\begin{table}[h!]
\centering
  \begin{tabular}{|c|c|c|}
    \hline
   Energy & $q\neq 0, \Lambda \neq 0, r\rightarrow\infty$ & $q=0, \Lambda \neq 0, r\rightarrow\infty$\\
    \hline
    $E_E$ & $\infty$ & $\infty$  \\
    $E_M$ & $-\infty$ &  $-\infty$ \\
    \hline
  \end{tabular}
  \caption{Limiting behaviour of the energy of the massless black hole solution in the Einstein and M\o ller prescriptions.}
\end{table}

According to the obtained results, we come to the conclusion that the Einstein and M\o ller energy-momentum complexes may provide an instructive tool for the study of the energy-momentum localization of gravitating systems, although in this work we cannot reach a definite answer to the problem of localization. That being said,  the investigation of the problem of the energy-momentum localization in the context of the black hole solution considered here, by applying other energy-momentum complexes as well as the teleparallel equivalent of general relativity (TEGR), is planned as a future perspective.

\section*{Acknowledgments}

The authors are grateful to the anonymous referees for their valuable comments and suggestions.\\
Farook Rahaman and Surajit Chattopadhyay are grateful to the Inter-University Centre for Astronomy and Astrophysics
(IUCAA), India, for providing the Associateship Programme. Farook Rahaman is thankful 
to DST, Govt. of India, for providing financial support under the SERB programme. Irina Radinschi is indebted to Prof. Rodica Tudorache and Prof. Dorel Fetcu from the Department of Mathematics and Informatics of the ``Gheorghe Asachi" Technical University, Iasi, Romania for their invaluable assistance.

\end{document}